\documentclass{ecai} 

\usepackage{latexsym}
\usepackage{amssymb}
\usepackage{amsmath}
\usepackage{amsthm}
\usepackage{booktabs}
\usepackage{enumitem}
\usepackage{graphicx}
\usepackage{color}

\newcommand{\BibTeX}{B\kern-.05em{\sc i\kern-.025em b}\kern-.08em\TeX}

\begin{document}


\begin{frontmatter}


\paperid{207} 


\title{Revocable Backdoor for Deep Model Trading}


\author[A]{\fnms{Yiran}~\snm{Xu}\footnote{Equal contribution.}}
\author[A]{\fnms{Nan}~\snm{Zhong}\footnotemark}
\author[A]{\fnms{Zhenxing}~\snm{Qian}\thanks{Corresponding Author. Email: zxqian@fudan.edu.cn}} 
\author[A]{\fnms{Xinpeng}~\snm{Zhang}\thanks{Corresponding Author. Email: zhangxinpeng@fudan.edu.cn}} 

\address[A]{Fudan University}


\begin{abstract}
Deep models are being applied in numerous fields and have become a new important digital product. Meanwhile, previous studies have shown that deep models are vulnerable to backdoor attacks, in which compromised models return attacker-desired results when a trigger appears. Backdoor attacks severely break the trustworthiness of deep models. In this paper, we turn this weakness of deep models into a strength, and propose a novel revocable backdoor and deep model trading scenario. Specifically, we aim to compromise deep models without degrading their performance, meanwhile, we can easily detoxify poisoned models without re-training the models. We design specific mask matrices to manage the internal feature maps of the models. These mask matrices can be used to deactivate the backdoors. The revocable backdoor can be adopted in the deep model trading scenario. Sellers train models with revocable backdoors as a trial version. Buyers pay a deposit to sellers and obtain a trial version of the deep model. If buyers are satisfied with the trial version, they pay a final payment to sellers and sellers send mask matrices to buyers to withdraw revocable backdoors. We demonstrate the feasibility and robustness of our revocable backdoor by various datasets and network architectures.
\end{abstract}

\end{frontmatter}


\section{Introduction}

Deep learning has achieved impressive performance in a wide range of fields \cite{he2017mask,ren2015faster,liu2016recurrent}. However, the security vulnerability of deep models, such as the notorious backdoor attack, hinders the deployment of deep models in some risk-sensitive domains \cite{badue2021self, du2022elements}. Most attackers \cite{gu2017badnets, li2021invisible} implant backdoors into a clean model during the training phase by data poisoning. The compromised models behave normally during the evaluation phase, whereas they return attacker-desired results when the predefined trigger appears. Considering the stealthiness and hazard of backdoor attacks, it has attracted a lot of attention in the machine learning security community \cite{nguyen2020input,wang2022training,hong2022handcrafted,zhanginject}.

Although backdoor attack is seen as a security threat or vulnerability for deep models in most studies, some researchers adopt backdoor attack to conduct some positive purposes \cite{adi2018turning,shan2020gotta,lin2021you}. There are various positive usages of the backdoor like dataset or model copyright protection \cite{adi2018turning}, artificial intelligence interpretability \cite{lin2021you}, adversarial example defence \cite{shan2020gotta}, etc. Adi et al. \cite{adi2018turning} adopt the backdoor attack to implement model watermarking for model copyright protection. They use significantly abnormal outputs of models triggered by poisonous inputs as the watermarking signal to verify the ownership of the model. A similar idea \cite{li2022untargeted} is also implemented to achieve dataset ownership protection. Shan et al. \cite{shan2020gotta} utilize the backdoor as a honeypot for deep models to resist adversarial example attack. When an attacker constructs an adversarial example for the backdoored model, the distribution of the interior feature map of the adversarial example is close to that of poisonous ones. Therefore, defenders identify adversarial example input by inspecting the distribution of the interior feature map.

\begin{figure*}[t]
   \centering
   \includegraphics[width=6.5in]{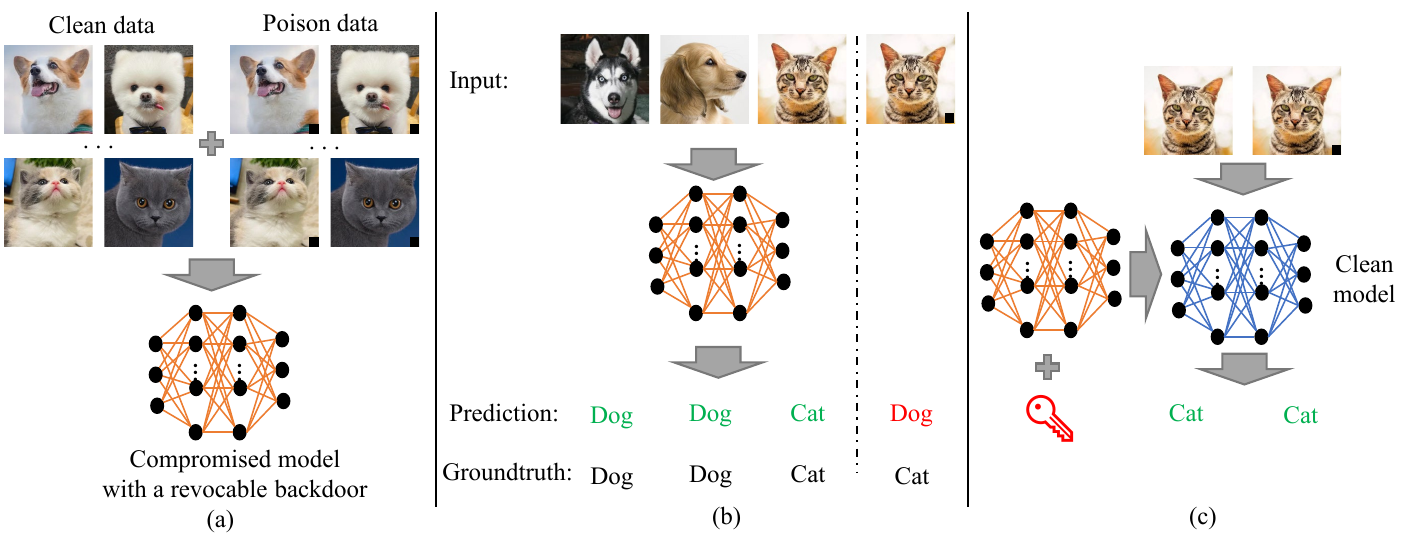}
   \caption{The illustration of the revocable backdoor for the model trading scenario. (\textbf{a}) Seller trains a model with a revocable backdoor as a trial version for buyers. The performance of the trial version over clean inputs is the same as the final model (or a clean model without backdoors). (\textbf{b}) Buyers pay a deposit and obtain a trial model. Then, buyers evaluate whether the model meets their requirements. For trial models (backdoored models), the trigger predefined by sellers in (\textbf{a}) can lead models to return wrong results. Note that we use a pure small black square to represent the trigger pattern in the figure for simplicity. The practical trigger pattern adopted in our approach is more sophisticated. (\textbf{c}) if buyers decide to pay the final payment, sellers withdraw the hidden backdoors. The deep model returns the correct result even the trigger pattern appears. }
   \label{fig_intro}
\end{figure*}

In this paper, we first propose a novel concept of the revocable backdoor and explore its promising application which is deep model trading. The revocable backdoor denotes that attackers implant backdoors into clean deep models while attackers can easily withdraw these backdoors. In other words, attackers control poisoning and detoxification for deep models at the same time. The detoxification should be easy and efficient. Revocable backdoor still needs to satisfy the general properties of a backdoor. When inputs contain the trigger, the model with revocable backdoor produces the attacker's predefined outputs. Otherwise, the model should behave normally and produce correct output. Additionally, it should possess stealth and robustness to prevent users from detecting and removing the backdoor. The revocable backdoor can be applied to deep model trading to protect the rights and interests of both buyers and sellers.

The main contributions of this paper are three-fold: 
\begin{enumerate}
    \item We propose a promising deep model trading scenario, and hold that the revocable backdoor can address the issues haunted by buyers and sellers. 
    \item We propose a practical revocable backdoor, which does not require extra clean or poison data to withdraw backdoors. 
    \item Extensive experiments demonstrate that our proposed revocable backdoor is feasible and robust. 
\end{enumerate}

\section{Model trading Scenario}

Since the wild usage of deep learning models in real world, models become a tradable products. However, how to sale a trained model or buy a trained model still remains a question. The Uniform Commercial Code (UCC) \cite{white2022uniform} stipulates that when the goods delivered by the seller do not conform to the contract, the buyer has the significant right to "revoke acceptance". Correspondingly, when the buyer breaches the contract, such as improperly rejecting the goods, incorrectly revoking the acceptance of the goods, failing to pay the price etc., the seller has the right to undertake a series of remedial measures, including recovering the goods, reselling, and so on.

Therefore, the rights and demands of both the buyer and seller in the model transaction can be summarized as follows:

\textbf{Sellers} seek to profit by selling their own models. As models are considered data resources, a model "return" would carry immeasurable loss of benefits. They attempt to control costs as much as possible to increase profits.

\textbf{Buyers} hope to purchase models that meet their needs. When they find that the purchased model does not meet their needs, they want to be able to return it and get a refund.

\textbf{Trading platform} provides a reliable platform for model transactions, safeguarding the reasonable rights and interests of both buyers and sellers.

However, there are still a few issues:
\begin{itemize}
    \item \textbf{Q1}. How to allow the buyer to return and refund the purchased model when they find it does not meet their needs, ensuring their right to "revoke acceptance"?
    
    \item \textbf{Q2}. How to protect the seller's interests without being damaged when the buyer returns the model? That is, how to prevent the buyer from saving some of the data after receiving the model, and then returning and refunding it.
\end{itemize}

A possible solution to these questions could be to send a trial version at the time of delivery. The full version would then be delivered once the buyer confirms receipt. Then, the questions turn into:

\begin{itemize}
    \item \textbf{Q3}. How to ensure that both the trial version and the full version of the model meet the buyer's requirements?
    
    \item \textbf{Q4}. How to ensure the model will not be resold by the buyer?
\end{itemize}

To solve these two questions, we propose the concept of a revocable backdoor. Fig. \ref{fig_intro} illustrates the details of the deep model trading scenario. A buyer distributes his requirement to a seller. Then, the seller trains a model based on the buyer's requirements. The model has a revocable backdoor and can be sent to the buyer as a trail version. The buyer pays a deposit to obtain the trial model whose performance is almost the same as the clean final model over a normal task. Next, the buyer evaluates the trial model on his/her application and determines whether accepts the model and pays a final payment. If the buyer is satisfied with the trial version, they can pay the seller the final balance. After that, the seller sends the mask matrix to the buyer and the revocable backdoor can be removed by the mask matrix. 

 Buyers can disclose the backdoor hidden in the model publicly and make the trial model unusable. Fig. \ref{fig_intro} illustrates the details of the deep model trading scenario.
The advantages of using the revocable backdoor are as follows:

\begin{enumerate}
    \item It allows the user to try the model while securing the rights of the seller. In the worst case, the buyer maliciously refuses to pay the final payment and deploys the trial model in his/her application. The seller can disclose the backdoor hidden in the model publicly and makes the trail model unusable. Or, if there is a third part (sale platform) involved, the seller can register the mask and sue the buyer for infringement. (Address \textbf{Q1} and \textbf{Q2}.)
    
    \item It also ensures the relevance between the trial version and the complete version of the model, preventing the buyer from receiving a complete model that does not match the trial version after paying the balance and confirming receipt. (Address \textbf{Q3}.)
    
    \item There is no backdoor in the complete version of the model, but the seller can still use the mask matrix to identify whether a model is the one they sold (like a serial number of a software). That is, by subtracting the mask matrix from the model, determining whether the model with the mask matrix removed contains a preset backdoor to authenticate the model's genuineness. (Address \textbf{Q4}.)
\end{enumerate}


\section{Related Work}

\subsection{Backdoor Attack}
Backdoor attack is an emerging topic in the machine learning security community \cite{li2022backdoor}. The goal of the backdoor attack is to compromise deep models, which return normal results for clean inputs, but return attacker-desired results when the trigger appears. Gu et al. \cite{gu2017badnets} first propose the concept and specific practical backdoor attack approach named BadNets. BadNets mounts an attack through data poison during the training phase. A conspicuous colourful square is adopted as a trigger pattern. BadNets generate some poisonous images by adding the trigger pattern to some clean images and changing their label to the target (attacker-desired) label. Such poisonous images are mixed with other clean training sets. The deep model trained over a poisonous training set contains an insidious backdoor which only activated by the trigger pattern. The implementation of the backdoor attack can be roughly grouped into two categories. 

\textbf{Poisoning-based backdoor} mounts an attack by inserting some well-designed poisonous samples into the training set \cite{li2020invisible, li2021invisible, nguyen2020wanet, chen2017targeted, guo2023physical}. Previous researchers focus on designing stealthy and efficient trigger patterns. Although BadNets proposes a seminal approach for backdoor attacks, its trigger pattern is conspicuous. In the follow-up studies, researchers explore some invisible trigger patterns. Li et al. \cite{li2020invisible} and Li et al. \cite{li2021invisible} adopt image steganographic noise ( Least Significant Bit and DNNs-based steganographic ) as a trigger pattern. The visual quality of poisonous is greatly boosted by invisible trigger patterns. Furthermore, WaNet \cite{nguyen2020wanet} adopts image affine transformation as a trigger pattern. The poisonous images can be viewed as slightly warped from clean ones. WaNet works effectively not only in the digital domain but also attacks successfully in the physical domain.

\textbf{Training process controlling-based backdoor} can achieve more stealthy attacks than poisoning-based ones \cite{nzhong_ijcai, wang2022bppattack,doan2021backdoor,doan2021lira}. As a trade-off, these approaches entail controlling the entire training phase. These attacks focus on both the stealthiness of the trigger pattern and compromised models. Backdoor attacks may leave traces in compromised models. Doan et al. \cite{doan2021backdoor} focus on the feature map separability of compromised models and employ Wasserstein distance to minimize the feature map distance between poisonous and clean inputs. Some similar ideas are proposed in subsequent studies. Besides consideration of the feature map domain, BppAttack \cite{wang2022bppattack} employs contrastive supervised learning and adversarial training instead of standard cross-entropy loss in the classification task. This learning approach ensures that compromised models steadily and successfully converge.


\subsection{Backdoor Defence}
Considering the serious threat of backdoor attacks to machine learning security, backdoor defence, which is the countermeasure against backdoor attacks, rapidly develops in recent studies \cite{zengadversarial,chen2022quarantine,liu2022complex,huangbackdoor}. There are two main categories in the backdoor defence, i.e., backdoor detection \cite{wang2019neural,guo2021aeva}, backdoor purification \cite{wu2021adversarial,liu2018fine,li2020neural,zheng2022data}
In terms of backdoor detection approaches, they aim to determine whether the deep model contains a backdoor. Neural Cleanse (NC) \cite{wang2019neural} is one of the well-known backdoor detection algorithms. NC observes that the backdoor of the classifier constructs a shortcut between the target label and other clean labels. Based on this observation, NC employs a abnormality detection algorithm to determine whether the model is poisoned by analyzing the size of perturbation leading all clean images to the target label. 

Compared with backdoor detection, backdoor purification aims to eliminate backdoors on the premise that the performance of clean inputs does not significantly degrade. Fine-Pruning \cite{liu2018fine} adopts network pruning to erase backdoors hidden in the models. For existing neural networks, most neurons are dormant when the clean inputs come. However, these dormant neurons may be activated by the trigger pattern and lead models to return attacker-desired results. Therefore, Fine-Pruning identifies dormant neurons based on some clean inputs then deletes them and fine-tunes the model. NAD (Neural Attention Distillation) \cite{li2020neural} is an alternative effective backdoor purification approach, which utilizes model distillation to erase backdoors. NAD fine-tunes backdoored model to generate a teacher model which is used to conduct model distillation. Many other sophisticated model purification approaches are proposed in subsequent studies.



\section{Revocable Backdoor}

\begin{figure*}[t]
   \centering
   \includegraphics[width=6.8in]{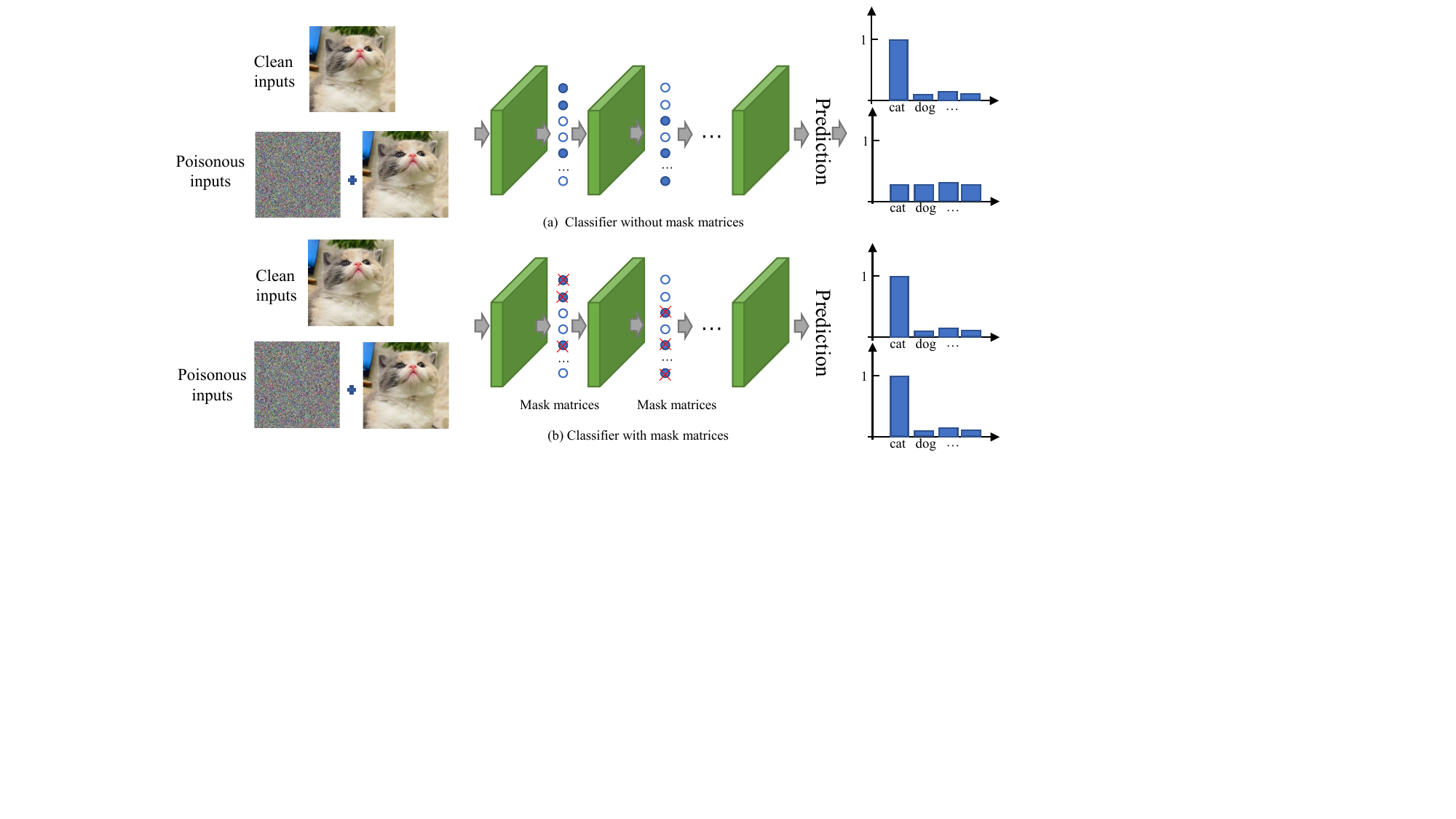}
   \caption{The framework of the implementation revocable backdoor attack. Our approach consists of two main parts. (a) is similar to common backdoor attacks. The training set including both clean and poisonous inputs is fed into the classifier, which returns correct results for clean inputs yet wrong results for poisonous ones. (b) is the crux of the revocability of our backdoor. We withdraw our backdoor by controlling the interior feature map. We utilize some trainable mask matrices to intentionally break the poisonous inference link. }
   \label{fig_framework}
\end{figure*}
 
\subsection{Preliminaries}

In this paper, we focus on the image classification task which is consistent with previous studies. Since the revocable backdoor is first proposed by ours, we describe more details of the revocable backdoor requirement applied in the deep model trading. 

\textbf{Goal} The revocable backdoor aims to ensure that buyers complete the final payment when they are satisfied with the trial model. If buyers refuse to complete the final payment, sellers can disclose the hidden backdoor and corresponding trigger pattern. For example, buyers obtain a face recognition model with a revocable backdoor and deploy this model in their company. If the backdoor is disclosed, anyone with the trigger pattern can break the face recognition model, that is, the model is unusable. Note that our revocable backdoor does not aim to protect the copyright of the model, which is usually accomplished by model watermarking \cite{adi2018turning}. Conventional backdoor-based watermarking aims to verify the ownership of the model, whose goal is completely different from ours. Although backdoor-based watermarking also can make the model return incorrect results when the trigger appears, the backdoor is permanent and irrevocable. Even though buyers complete the final payment, the backdoor still exists in the model and brings security threats to the model deployment.

\textbf{Effectiveness} In previous backdoor attacks, attackers aim to make the compromised classifier return the target label when the trigger appears \cite{gu2017badnets, nguyen2020wanet}. However, in the context of model tradings, the goal of our revocable backdoor for model sellers is different from previous studies. Sellers can distribute the trigger pattern to make the classifier held by buyers unusable in the event that the buyer does not complete the final payment. Therefore, our revocable backdoor aims to decrease the classification accuracy as low as possible when the trigger appears. In other words, we view the attack success if the compromised classifier returns an arbitrarily wrong label. 

\textbf{Fidelity} The classifier with a revocable backdoor can be used as a trial version of the final classifier for buyers. Therefore, the revocable backdoor aims to keep the performance of the backdoored classifier over clean inputs the same as a clean classifier. For buyers, the performance of the trial version over clean inputs is almost identical to the final version (backdoor-free). In this case, buyers can evaluate whether the classifier meets their requirements.

\textbf{Revocability} The crux of our revocable backdoor is that we can easily withdraw the backdoor without requiring an extra training phase or clean/poisonous data. The revocability denotes that the compromised classifier whose backdoor is withdrawn can correctly predict poisonous inputs.

\textbf{Robustness} In the model trading scenario, the backdoors hidden in the compromised model can and only can be withdrawn (erased) by sellers. Therefore, our attack should ensure that the backdoor is resistant to various backdoor purification defences.

\subsection{Method}\label{sub_method}

Fig. \ref{fig_framework} illustrates the framework of our approach. 
Let $\mathcal{D} = \{ ({x}_i, y_i) \}_{i=1}^{N}$ denotes the benign training set, where ${x}_i \in \mathcal{X}= \{0,1,\ldots, 255\}^{C\times W \times H}$ is the image, $y_i \in \mathcal{Y} = \{1,\ldots, N\}$ is its label, and $N$ is the number of classes. The trigger pattern is denoted as $\delta$, where $\delta \in \mathcal{P}=\{-t,-t+1,\ldots,t+1,t\}^{C\times W \times H}$. The elements of the trigger pattern are between $-t$ and $t$. In the backdoor attack, we select a portion of clean images ${x}$, which are stamped with the trigger pattern, as poisonous ones. We denote the clean training set and poisonous training set as $\mathcal{D}_c$ and $\mathcal{D}_p$, respectively. Given a deep neural network-based classifier $f_{\theta}(\cdot)$, we aim to minimize the following equation,
\begin{eqnarray}
   L_{bd} = \frac{1}{{|D_c|}}\sum\limits_{(x,y) \in D_c} \ell  \left( {{f_\theta }(x),y} \right) \\ - \frac{1}{{|D_p|}}\sum\limits_{(x,y) \in D_p} max( \ell   \left( {{f_\theta }(x+\delta),y),c} \right), \label{e1}
\end{eqnarray}
where $\ell(\cdot)$ denotes cross-entropy loss widely adopted in classification tasks. We update $f_\theta(\cdot)$ and $\delta$ to minimize equation \ref{e1}. The core idea of equation \ref{e1} is to achieve the effectiveness and fidelity of the revocable backdoor attack. The first term of equation \ref{e1} ensures that the performance of backdoored classifiers over clean inputs does not drop. The second term denotes maximizing the cross-entropy loss of poisonous inputs, i.e., making the classifier return wrong results when the trigger appears. However, maximizing the cross-entropy loss of poisonous inputs is much easier than minimizing the cross-entropy loss of clean inputs. In other words, classifier $f_\theta(\cdot)$ easily learns how to give wrong results for poisonous inputs but it is hard to learn how to give correct results for clean inputs. As a result, we adopt $max(\cdot)$  and hyperparameter confidence $c$ to restrict the second term. For a $N$ categories classification task, if the cross-entropy loss of input is larger than $-log(1/N)$, the classifier will return an incorrect result. Therefore, we set $c=-1.1\times log(1/N)$ that is slightly larger than the minimum correct decision threshold for a $N$ categories classifier. If we remove the restriction of $max(\cdot)$ or set an excessively large $c$, the classifier $f_\theta$ will only learn the second term of equation \ref{e1} and overlook the first term.

Besides $L_{bd}$, our approach also entails considering how to withdraw our backdoor. The modern deep neural network-based classifier $f_{\theta}(\cdot)$ is made up of some cascade layers. The output of $f_{\theta}(\cdot)$ can be expressed as,
\begin{eqnarray}
   f_\theta(\cdot) = f_\mathrm{n}\left(\mathrm{\cdots}f_\mathrm{2}\mathrm{(} f_\mathrm{1}{(\cdots))} \right),
\end{eqnarray}
where $n$ denotes the number of layers. The output of the interior layer is named as feature map. Based on the cascade structure of $f_{\theta}(\cdot)$, we employ some matrices named masks to manipulate feature maps. Compromised classifiers learn two tasks, i.e., a normal task and an insidious backdoor task. These two tasks conduct different inference links. In other words, the feature maps of these two tasks are different. Therefore, we can adopt some mask matrices to break backdoors without affecting clean inputs. We denote the classifier $f_{\theta}(\cdot)$ with mask matrices $M_{k}$ as $f_{\theta_{mask}}(\cdot)$. The formulaic expression is shown as follows,
\begin{eqnarray}
 f_{\theta_{mask}}(\cdot) =  f_\mathrm{n}\left(\mathrm{\cdots} M_2\cdot f_\mathrm{2}\mathrm{(} M_1\cdot f_\mathrm{1}{(x))} \right),
\end{eqnarray}
where $M$ and $\cdot$ denote trainable mask matrices and Hadamard product, respectively. The size of $M_{k}$ is the same as the corresponding feature map. The initial elements of $M_{k}$ are filled with 1. Moreover, we require a regularization term for $M_{k}$. The goal of regularization is used to make the elements of $M_{k}$ as close to 1 as possible. The regularization for $M_{k}$ is expressed as follows,

\begin{eqnarray}
L_{R} = sum(|M_{k_{ij}} - 1 \vert ),
\end{eqnarray}
where $sum(\cdot)$ denotes the sum of all elements.
The usage of $L_{R}$ forces $M_{k}$ to change the most key elements related to poisonous inputs in feature maps. $L_{R}$ can effectively mitigate the impact of $M_{k}$ on clean inputs. For $f_{\theta_{mask}}(\cdot)$, it should return correct results regardless of whether the trigger pattern $\delta$ appears. Therefore, the loss function for $f_{\theta_{mask}}(\cdot)$ is expressed as follows,
\begin{eqnarray}
   L_{rev} = [\frac{1}{{|D_c|}}\sum\limits_{(x,y) \in D_c} \ell  \left( {{f_{\theta_{mask}} }(x),y} \right) \\ + \frac{1}{{|D_p|}}\sum\limits_{(x,y) \in D_p} \ell  \left( {{f_{\theta_{mask}} }(x+\delta),y} \right)]. \label{e_rev}
\end{eqnarray}
We update $M_{k}$ by minimizing $L_{rev}$ and make $M_{k}$ adaptive find and break the poisonous inference link.
The final cost function is expressed as follows,
\begin{eqnarray}
L = L_{bd}+L_{rev}+\alpha\cdot L_{R},\label{e_tot}
\end{eqnarray}
where $\alpha$ is the balance factor to control the weight of the regularization. We simultaneously update parameters of $f_{\theta}$, trigger pattern $\delta$, and mask matrices $M_{k}$ to minimize $L$. The backdoor hidden in the well-trained $f_{\theta}$ can easily be withdrawn by adding $M_{k}$ in the interior layers.  $f_{\theta}(\cdot)$ and $f_{\theta_{mask}}(\cdot)$ can be used as the trial model and final model for sellers during the model trading.

\subsection{Trigger Fine-tuning} \label{ft_trigger}
In the previous subsection \ref{sub_method}, we describe our approach to create a compromised classifier whose backdoors are controlled by mask matrices. The trigger pattern $\delta$, whose elements are between $-t$ and $t$, is trained with the classifier at the same time. In this subsection, we delve into the design of the trigger pattern to control the balance of the trigger pattern imperceptibility and attack robustness. Hyperparameter $t$ controls the modification magnitude of the trigger pattern. In other words, a small $t$ ensures the imperceptibility of the trigger pattern. In this subsection, we frozen classifier $f_\theta$ and mask matrices $M$, i.e., $f_{\theta_{mask}}$ is also frozen, and update trigger patterns. The loss function of updating trigger $\delta$ is as follows,
\begin{eqnarray}
L = L_{bd}+L_{rev}+\beta\cdot | \delta \vert_2,
\end{eqnarray}
where $| \delta \vert_2$ denotes the $L_2$-norm of the trigger $\delta$, and $\beta$ controls the balance of $| \delta \vert$. We control the modification magnitude of the trigger pattern $\delta$ by changing the value of the hyperparameter $\beta$ and $t$. 

\subsection{Backdoor Erasing for Existing Attacks}
To the best of our knowledge, we are the first to propose the concept of the revocable backdoor. For existing conventional backdoor attacks like BadNtes \cite{gu2017badnets}, WaNet \cite{nguyen2020wanet}, Blend \cite{chen2017targeted}, etc. They can not be withdrawn by attackers. In this subsection, we describe a backdoor erasing method to withdraw existing backdoors. In the model trading scenario, we suppose that buyers own some clean images and their corresponding poisonous ones. Then, they update the parameters of classifier $f_\theta$ by minimizing the following loss as,
\begin{eqnarray}
   L_{erasing} = [\frac{1}{{|D_c|}}\sum\limits_{(x,y) \in D_c} \ell  \left( {{f_\theta }(x),y} \right) \\ + \frac{1}{{|D_p|}}\sum\limits_{(x,y) \in D_p} \ell  \left( {{f_\theta }(x+\delta),y} \right)]. \label{unlearning}
\end{eqnarray}
Compared with equation \ref{e1}, $L_{erasing}$ minimizes the cross-entropy loss between clean and poisonous images and their label. The number of poisonous images is critical to the performance of backdoor erasing. We employ this method as our baseline. Compared with our proposed method \ref{sub_method} and \ref{ft_trigger}, this method requires gradient calculation and some poisonous images, which is not practical in the model trading scenario.  


\section{Experimental Results}

\begin{table*}[]
  \centering

    \begin{tabular}{cccccc}
    \toprule
    Dataset &       & \multicolumn{2}{c}{Effectiveness } & \multicolumn{2}{c}{Revocability} \\
    $Acc_{Clean}$ & Attack & $Acc_{BD}$-C ($\uparrow$) & $Acc_{BD}$-P ($\downarrow$) & $Acc_{BD}$-C ($\uparrow$) & $Acc_{BD}$-P ($\uparrow$) \\
    \hline
          & BadNets & 82.99  & 10.04  & 79.90  & 79.68  \\
          & Blend & 83.45  & 10.03  & 78.49  & 67.51  \\
    CIFAR-10  & SIG   & 83.73  & 10.00  & 78.57  & 77.57  \\
    85.17 & LSB   & 77.58  & 77.54  & 75.28  & 75.12  \\
          & WaNet & 79.30  & 31.90  & 77.77  & 77.34  \\
          & BppAttack & 77.85  & 77.32  & 74.65  & 74.71  \\
          & Ours  & \textbf{85.88}  & \textbf{0.57}  & \textbf{85.90}  & \textbf{84.77}  \\
          \hline
          & BadNets & 97.43  & 4.99  & 96.79  & 96.78  \\
          & Blend & 97.10  & 0.15  & 94.06  & 93.48  \\
    GTSRB & SIG   & 97.38  & \textbf{0.02}  & 95.62  & 95.15  \\
    98.38 & LSB   & 97.87  & 4.08  & 96.77  & 96.77  \\
          & WaNet & 95.83  & 6.12  & 95.78  & 96.27  \\
          & BppAttack & 95.70  & 95.24  & 96.71  & 96.68  \\
          & Ours  & \textbf{97.69}  & 0.12  & \textbf{97.95}  & \textbf{97.66}  \\
          \hline
          & BadNets & 63.26  & 63.29  & 64.74  & 64.60  \\
          & Blend & 70.52  & 9.05  & 64.50  & 61.07  \\
    Sub-ImageNet & SIG   & 67.87  & \textbf{8.92}  & 67.53  & 66.39  \\
    73.91 & LSB   & 62.25  & 62.35  & 63.26  & 63.36  \\
          & WaNet & 64.10  & 63.29  & 65.01  & 64.91  \\
          & BppAttack & 63.90  & 63.73  & 64.67  & 64.67  \\
          & Ours  & \textbf{74.06}  & 10.43  & \textbf{74.33}  & \textbf{71.33}  \\
          \bottomrule
    \end{tabular}%
      \caption{The experimental comparison (ResNet-18) between baselines and ours. $Acc_{Clean}$, $Acc_{BD}$-C, and $Acc_{BD}$-P denote the accuracy of the clean classifier, the accuracy of the compromised classifier for clean inputs, and the accuracy of the compromised classifier for poisonous inputs, respectively. Effectiveness and revocability denote the compromised classifier and compromised classifier with mask matrices, respectively. }
  \label{tab_main}%
\end{table*}%

\subsection{Experimental Setup}

\textbf{Datasets and Networks} In this paper, we adopt three datasets including CIFAR-10 \cite{krizhevsky2009learning}, GTSRB \cite{stallkamp2012man}, and Sub-ImageNet to evaluate the performance of our approach. These datasets are widely used in previous studies \cite{nguyen2020wanet, wang2022bppattack}. The number of categories of CIFAR-10 and GTSRB is 10, and 43, respectively. Sub-ImageNet consists of 10 categories which are randomly selected from the original ImageNet dataset \cite{deng2009imagenet}. The image size of CIFAR-10, GTSRB, and Sub-ImageNet is $32\times 32$,  $32\times 32$, and $224\times 224$, respectively. In terms of network architectures, we adopt ResNet-18 \cite{he2016deep} and VGG \cite{simonyan2014very} in the experiments.  

\textbf{Backdoor Baselines} Since we are the first to propose the concept of the revocable backdoor attack, there are no suitable revocable backdoor attacks used as a comparison. We suppose that users obtain some clean and corresponding poisonous images to finetune to a backdoored classifier to achieve revocability based on existing conventional attacks. We adopt various trigger patterns to conduct backdoor attacks. The trigger pattern includes BadNets \cite{gu2017badnets}, Blend \cite{chen2017targeted}, SIG \cite{barni2019new}, LSB \cite{li2020invisible}, WaNet \cite{nguyen2020wanet}, and BppAttack \cite{wang2022bppattack}. The details setting of baseline triggers can be found in the supplement. The compromised (backdoored) classifier is generated by updating equation \ref{e1} for baselines. The poisoning rate is fixed as $r=0.1$. In the backdoor erasing phase, we suppose that users obtain $o=5$\% clean data of the training set and corresponding poisonous ones. Then, we finetune the backdoored baseline classifiers with equation \ref{unlearning}.

\textbf{Implement Details}
Our approach consists of two steps, training revocable backdoored classifier and finetuning trigger pattern. The training epoch and batch size are 180, and 256, respectively. We adopt SGD (Stochastic Gradient Descent) optimizer for updating the parameters of the classifier. The initial learning rate of SGD is 0.01. For trainable mask matrices and trigger pattern $\delta$, we adopt Adam optimizer with a 0.001 initial learning rate. The poisoning rate is 0.5 for the first 80 epochs. The poisoning rate times by 0.5 in intervals of 10 epochs when the iteration epoch exceeds 80. Hyperparameter $\alpha$ is fixed as 10, 10, and 100 for CIFAR-10, GTSRB, and Sub-ImageNet, respectively. In the second phase, we fixed the poisoning rate as 0.5. The hyperparameter $\beta$ is 0.01 for all datasets. The hyperparameter $t$ controlling the range of our trigger pattern is 10 in both phases.


\subsection{Attack Effectiveness}
We adopt accuracy to measure the performance of the classifier. In this part, we evaluate the Effectiveness, Fidelity and Revocability of our approach. The results of our approach and baselines are shown in Table \ref{tab_main}. We also train a clean classifier whose training setup is the same as compromised ones as a reference.  We aim to significantly degrade the classifier accuracy when the trigger appears while keeping the accuracy for clean inputs unchanged. In terms of revocability, the goal of mask matrices employed in our attack is to withdraw the backdoor. From Table \ref{tab_main}, our approach achieves satisfactory results in the aspect of attack effectiveness, fidelity, and revocability. The accuracy for clean inputs is almost unchanged compared with the clean classifier in most cases. The accuracy for poisonous inputs is decreased to near even less than random guesses, meanwhile, it recovers to standard performance (like a clean classifier) when we insert mask matrices to withdraw the backdoor. For baselines, their performance is significantly inferior to ours. In previous common backdoor studies \cite{gu2017badnets,chen2017targeted,nguyen2020wanet,wang2022bppattack}, their attack setting is that a compromised classifier returns the target label when the trigger appears. This attack setting is much easier than our setting, in which the compromised classifier does not need to understand the semantic information of inputs and only builds a strong relation between the trigger and target label. However, the compromised need to understand the semantic information of inputs and then return wrong results when the trigger appears in our attack setting. Therefore, some invisible triggers (like LSB and BppAttack) whose perturbations are small cannot successfully compromise the classifier. The experiments of VGG network can be found in Table 2 of the supplementary material.

\begin{figure*}[]
   \centering
   \includegraphics[width=6.8in]{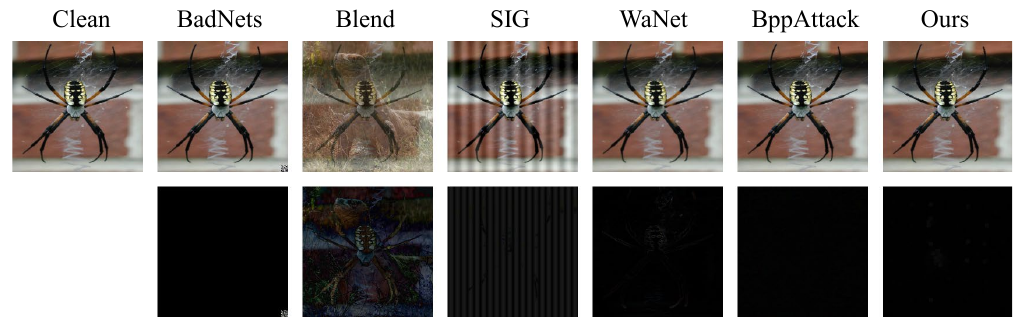}
   \caption{The visualization results of the trigger (Sub-ImgeNet). The first line denotes clean and poisonous images. The second line denotes the residual between clean and poisonous images. The visualization results of CIFAR-10 and GTSRB can be found in Figure 1 of the supplementary material.}
   \label{fig_visualization}
\end{figure*}

Apart from quantitative results, we also visualize our trigger pattern in Fig. \ref{fig_visualization}. Our approach achieves satisfactory visual results and users cannot perceive trigger patterns. Based on Fig. \ref{fig_visualization} and Table \ref{tab_main}, we find that only triggers (like BadNets) with large modification (visible) can attack successfully among baselines. Our approach achieve both trigger invisibility and attack effectiveness at the same time.  

\subsection{Attack Robustness}

We recall our revocable backdoor scenario, i.e., deep model trading scenario. For sellers, the goal of the revocable backdoor is to make the model unusable by distributing the trigger pattern to the public if buyers do not pay the final payment. Sellers can send the trial model to buyers and let them know the existence of the hidden backdoor. Therefore, we do not consider backdoor detection defence in the robustness evaluation. We focus on evaluating the resistance of our backdoor against backdoor purification defences \cite{li2020neural,liu2018fine}. In other words, we evaluate whether buyers can adopt backdoor purification to erase the backdoor without the permission of sellers. 

The easiest backdoor purification method, which may be adopted by buyers, is fine-tuning. Buyers collect some clean images whose distribution is similar to the training set to fine-tune the trial classifier containing our revocable backdoor. 
To simulate fine-tuning, we adopt SGD optimizer with 0.01 learning rate kept the same as the training phase to fine-tune the compromised classifier with 5\% clean images from the training set. We fine-tune the classifiers for 50 epochs. The experimental results have been shown in Table \ref{tab_NAD_FT}. Fine-tuning cannot remove our backdoor in all cases.

\begin{table*}[htbp]
  \centering
  
    \begin{tabular}{ccccccccc}
    \toprule
          & \multicolumn{2}{c}{No defence} & \multicolumn{2}{c}{NAD} & \multicolumn{2}{c}{Fine-tuning} & \multicolumn{2}{c}{Fine-pruning} \\
          \hline
    Dataset & $Acc_{BD}$-C & $Acc_{BD}$-P & $Acc_{BD}$-C & $Acc_{BD}$-P & $Acc_{BD}$-C & $Acc_{BD}$-P & $Acc_{BD}$-C & $Acc_{BD}$-P \\
    \hline
    CIFAR-10 & 85.88  & 0.57  & 86.16 & 3.64  & 86.05 & 0.53  &  84.88     &  6.92 \\
    GTSRB & 97.69  & 0.12  & 75.25 & 73.65 & 98.56 & 0.03  &   97.19    & 0.91  \\
    SubImageNet & 74.06  & 10.43  & 74.39 & 10.83 & 74.09 & 10.67 &  73.12     & 10.54 \\
     \bottomrule
    \end{tabular}%
    \caption{The experimental results of NAD defence, fine-tuning, and fine-pruning.}
  \label{tab_NAD_FT}%
\end{table*}%

An alternative well-known backdoor is Fine-pruning \cite{liu2018fine}, which combines network pruning and fine-tuning. As aforementioned related work part, most neurons are dormant for clean inputs. These dormant neurons almost do not affect the final performance of clean inputs. However, they may be activated by poisonous inputs. Therefore, Fine-pruning prunes dormant neurons with some clean inputs and fine-tunes the pruned classifier at the same time. The specific defence setup is as follows. We pruning the last layer of the classifier which is widely adopted in previous studies. The classifier is fine-tuned in the interval of 10 neurons. The optimizer is SGD with a 0.01 learning rate. We set the maximum tolerable drop of the clean task accuracy as less than 1\%. Table \ref{tab_NAD_FT} shows the results of Fine-pruning. It also cannot remove our backdoor. The details of the accuracy (both clean and poisonous inputs) with regard to the number of pruned neurons can be found in the supplementary material.

\begin{table}[]
  \centering
  
    \begin{tabular}{ccccc}
    \toprule
          & \multicolumn{2}{c}{w/o TF} & \multicolumn{2}{c}{ TF} \\
          \hline
    Dataset & PSNR  & SSIM  & PSNR  & SSIM \\
    CIFAR-10 & 34.15  & 0.96  & 35.63  & 0.97  \\
    GTSRB & 32.33  & 0.87  & 36.38  & 0.94  \\
    Sub-ImageNet & 32.27  & 0.73  & 44.45  & 0.95  \\
    \bottomrule
    \end{tabular}%
    \caption{The comparison of image quality between original and fine-tuned trigger. TF denotes trigger fine-tuning.}
  \label{tab_ablation_ft}%
\end{table}%

Besides fine-tuning and fine-pruning, we also evaluate the robustness of our approach by NAD \cite{li2020neural}. NAD first trains a teacher model. Then, NAD adopts the teacher model as a guide to conduct attention distillation. The experimental results of NAD are shown in Table \ref{tab_NAD_FT}. NAD is ineffective in our approach. Although NAD can erase the backdoor in some cases (GTSRB dataset), the accuracy of clean inputs severely degrades compared with the clean model.

\subsection{Ablation Studies}
In our approach, we adopt two steps to generate the trigger pattern. The second part, which fine-tunes the trigger, aims to control the perturbation of trigger pattern. Table \ref{tab_ablation_ft} gives the quantitative comparison between the trigger with or without fine-tuning. The PSNR (poisonous image quality) and SSIM \cite{wang2004image} is significantly improved benefit from the trigger fine-tuning.

More ablation studies over confidence $c$, regularization $L_R$ and the uniqueness of the mask matrices can be found in the supplement.

\section{Conclusions}

In this paper, we first propose the concept of the revocable backdoor attack and a novel application, that is, deep model trading. Revocable backdoor aims to create a trial model whose performance is similar to the clean/final model for buyers to evaluate its performance. Meanwhile, sellers can easily withdraw the backdoor hidden in the trial model when they obtain the final payment. The revocable backdoor attack is achieved by controlling the interior feature maps of models. We train models and mask matrices at the same time. The mask matrices are critical to turning the backdoored (trial) model into a clean (final) model. Extensive experiments demonstrate that our approach is feasible and robust.



\begin{ack}
This work was supported by the National Natural Science Foundation of China under Grants U20B2051, 62072114, U20A20178, U22B2047
\end{ack}



\bibliography{main}

\begin{thebibliography}{42}
\providecommand{\natexlab}[1]{#1}
\providecommand{\url}[1]{\texttt{#1}}
\expandafter\ifx\csname urlstyle\endcsname\relax
  \providecommand{\doi}[1]{doi: #1}\else
  \providecommand{\doi}{doi: \begingroup \urlstyle{rm}\Url}\fi

\bibitem[Adi et~al.(2018)Adi, Baum, Cisse, Pinkas, and Keshet]{adi2018turning}
Y.~Adi, C.~Baum, M.~Cisse, B.~Pinkas, and J.~Keshet.
\newblock Turning your weakness into a strength: Watermarking deep neural networks by backdooring.
\newblock In \emph{27th $\{$USENIX$\}$ Security Symposium ($\{$USENIX$\}$ Security 18)}, pages 1615--1631, 2018.

\bibitem[Badue et~al.(2021)Badue, Guidolini, Carneiro, Azevedo, Cardoso, Forechi, Jesus, Berriel, Paixao, Mutz, et~al.]{badue2021self}
C.~Badue, R.~Guidolini, R.~V. Carneiro, P.~Azevedo, V.~B. Cardoso, A.~Forechi, L.~Jesus, R.~Berriel, T.~M. Paixao, F.~Mutz, et~al.
\newblock Self-driving cars: A survey.
\newblock \emph{Expert Systems with Applications}, 165:\penalty0 113816, 2021.

\bibitem[Barni et~al.(2019)Barni, Kallas, and Tondi]{barni2019new}
M.~Barni, K.~Kallas, and B.~Tondi.
\newblock A new backdoor attack in cnns by training set corruption without label poisoning.
\newblock In \emph{2019 IEEE International Conference on Image Processing (ICIP)}, pages 101--105. IEEE, 2019.

\bibitem[Chen et~al.(2022)Chen, Zhang, Zhang, Chang, Liu, and Wang]{chen2022quarantine}
T.~Chen, Z.~Zhang, Y.~Zhang, S.~Chang, S.~Liu, and Z.~Wang.
\newblock Quarantine: Sparsity can uncover the trojan attack trigger for free.
\newblock In \emph{Proceedings of the IEEE/CVF Conference on Computer Vision and Pattern Recognition}, pages 598--609, 2022.

\bibitem[Chen et~al.(2017)Chen, Liu, Li, Lu, and Song]{chen2017targeted}
X.~Chen, C.~Liu, B.~Li, K.~Lu, and D.~Song.
\newblock Targeted backdoor attacks on deep learning systems using data poisoning.
\newblock \emph{arXiv preprint arXiv:1712.05526}, 2017.

\bibitem[Deng et~al.(2009)Deng, Dong, Socher, Li, Li, and Fei-Fei]{deng2009imagenet}
J.~Deng, W.~Dong, R.~Socher, L.-J. Li, K.~Li, and L.~Fei-Fei.
\newblock Imagenet: A large-scale hierarchical image database.
\newblock In \emph{2009 IEEE conference on computer vision and pattern recognition}, pages 248--255. Ieee, 2009.

\bibitem[Doan et~al.(2021{\natexlab{a}})Doan, Lao, and Li]{doan2021backdoor}
K.~Doan, Y.~Lao, and P.~Li.
\newblock Backdoor attack with imperceptible input and latent modification.
\newblock \emph{Advances in Neural Information Processing Systems}, 34:\penalty0 18944--18957, 2021{\natexlab{a}}.

\bibitem[Doan et~al.(2021{\natexlab{b}})Doan, Lao, Zhao, and Li]{doan2021lira}
K.~Doan, Y.~Lao, W.~Zhao, and P.~Li.
\newblock Lira: Learnable, imperceptible and robust backdoor attacks.
\newblock In \emph{Proceedings of the IEEE/CVF International Conference on Computer Vision}, pages 11966--11976, 2021{\natexlab{b}}.

\bibitem[Du et~al.(2022)Du, Shi, Zeng, Zhang, and Mei]{du2022elements}
H.~Du, H.~Shi, D.~Zeng, X.-P. Zhang, and T.~Mei.
\newblock The elements of end-to-end deep face recognition: A survey of recent advances.
\newblock \emph{ACM Computing Surveys (CSUR)}, 54\penalty0 (10s):\penalty0 1--42, 2022.

\bibitem[Gu et~al.(2017)Gu, Dolan-Gavitt, and Garg]{gu2017badnets}
T.~Gu, B.~Dolan-Gavitt, and S.~Garg.
\newblock Badnets: Identifying vulnerabilities in the machine learning model supply chain.
\newblock \emph{arXiv preprint arXiv:1708.06733}, 2017.

\bibitem[Guo et~al.(2021)Guo, Li, and Liu]{guo2021aeva}
J.~Guo, A.~Li, and C.~Liu.
\newblock Aeva: Black-box backdoor detection using adversarial extreme value analysis.
\newblock In \emph{International Conference on Learning Representations}, 2021.

\bibitem[Guo et~al.(2023)Guo, Zhong, Qian, and Zhang]{guo2023physical}
Y.~Guo, N.~Zhong, Z.~Qian, and X.~Zhang.
\newblock Physical invisible backdoor based on camera imaging.
\newblock In \emph{Proceedings of the 31st ACM International Conference on Multimedia}, pages 7817--7825, 2023.

\bibitem[He et~al.(2016)He, Zhang, Ren, and Sun]{he2016deep}
K.~He, X.~Zhang, S.~Ren, and J.~Sun.
\newblock Deep residual learning for image recognition.
\newblock In \emph{Proceedings of the IEEE conference on computer vision and pattern recognition}, pages 770--778, 2016.

\bibitem[He et~al.(2017)He, Gkioxari, Doll{\'{a}}r, and Girshick]{he2017mask}
K.~He, G.~Gkioxari, P.~Doll{\'{a}}r, and R.~B. Girshick.
\newblock Mask {R-CNN}.
\newblock In \emph{{IEEE} International Conference on Computer Vision, {ICCV} 2017, Venice, Italy, October 22-29, 2017}, pages 2980--2988, 2017.

\bibitem[Hong et~al.(2022)Hong, Carlini, and Kurakin]{hong2022handcrafted}
S.~Hong, N.~Carlini, and A.~Kurakin.
\newblock Handcrafted backdoors in deep neural networks.
\newblock \emph{Advances in Neural Information Processing Systems}, 35:\penalty0 8068--8080, 2022.

\bibitem[Huang et~al.()Huang, Li, Wu, Qin, and Ren]{huangbackdoor}
K.~Huang, Y.~Li, B.~Wu, Z.~Qin, and K.~Ren.
\newblock Backdoor defense via decoupling the training process.
\newblock In \emph{International Conference on Learning Representations, 2022}.

\bibitem[Krizhevsky et~al.(2009)Krizhevsky, Hinton, et~al.]{krizhevsky2009learning}
A.~Krizhevsky, G.~Hinton, et~al.
\newblock Learning multiple layers of features from tiny images.
\newblock 2009.

\bibitem[Li et~al.(2020{\natexlab{a}})Li, Xue, Zhao, Zhu, and Zhang]{li2020invisible}
S.~Li, M.~Xue, B.~Z.~H. Zhao, H.~Zhu, and X.~Zhang.
\newblock Invisible backdoor attacks on deep neural networks via steganography and regularization.
\newblock \emph{IEEE Transactions on Dependable and Secure Computing}, 18\penalty0 (5):\penalty0 2088--2105, 2020{\natexlab{a}}.

\bibitem[Li et~al.(2020{\natexlab{b}})Li, Lyu, Koren, Lyu, Li, and Ma]{li2020neural}
Y.~Li, X.~Lyu, N.~Koren, L.~Lyu, B.~Li, and X.~Ma.
\newblock Neural attention distillation: Erasing backdoor triggers from deep neural networks.
\newblock In \emph{International Conference on Learning Representations}, 2020{\natexlab{b}}.

\bibitem[Li et~al.(2021)Li, Li, Wu, Li, He, and Lyu]{li2021invisible}
Y.~Li, Y.~Li, B.~Wu, L.~Li, R.~He, and S.~Lyu.
\newblock Invisible backdoor attack with sample-specific triggers.
\newblock In \emph{Proceedings of the IEEE/CVF International Conference on Computer Vision}, pages 16463--16472, 2021.

\bibitem[Li et~al.(2022{\natexlab{a}})Li, Bai, Jiang, Yang, Xia, and Li]{li2022untargeted}
Y.~Li, Y.~Bai, Y.~Jiang, Y.~Yang, S.-T. Xia, and B.~Li.
\newblock Untargeted backdoor watermark: Towards harmless and stealthy dataset copyright protection.
\newblock \emph{Advances in Neural Information Processing Systems}, 35:\penalty0 13238--13250, 2022{\natexlab{a}}.

\bibitem[Li et~al.(2022{\natexlab{b}})Li, Jiang, Li, and Xia]{li2022backdoor}
Y.~Li, Y.~Jiang, Z.~Li, and S.-T. Xia.
\newblock Backdoor learning: A survey.
\newblock \emph{IEEE Transactions on Neural Networks and Learning Systems}, 2022{\natexlab{b}}.

\bibitem[Lin et~al.(2021)Lin, Lee, and Celik]{lin2021you}
Y.-S. Lin, W.-C. Lee, and Z.~B. Celik.
\newblock What do you see? evaluation of explainable artificial intelligence (xai) interpretability through neural backdoors.
\newblock In \emph{Proceedings of the 27th ACM SIGKDD Conference on Knowledge Discovery \& Data Mining}, pages 1027--1035, 2021.

\bibitem[Liu et~al.(2018)Liu, Dolan-Gavitt, and Garg]{liu2018fine}
K.~Liu, B.~Dolan-Gavitt, and S.~Garg.
\newblock Fine-pruning: Defending against backdooring attacks on deep neural networks.
\newblock In \emph{International Symposium on Research in Attacks, Intrusions, and Defenses}, pages 273--294. Springer, 2018.

\bibitem[Liu et~al.(2016)Liu, Qiu, and Huang]{liu2016recurrent}
P.~Liu, X.~Qiu, and X.~Huang.
\newblock Recurrent neural network for text classification with multi-task learning.
\newblock In \emph{Proceedings of the Twenty-Fifth International Joint Conference on Artificial Intelligence}, pages 2873--2879, 2016.

\bibitem[Liu et~al.(2022)Liu, Shen, Tao, Wang, Ma, and Zhang]{liu2022complex}
Y.~Liu, G.~Shen, G.~Tao, Z.~Wang, S.~Ma, and X.~Zhang.
\newblock Complex backdoor detection by symmetric feature differencing.
\newblock In \emph{Proceedings of the IEEE/CVF Conference on Computer Vision and Pattern Recognition}, pages 15003--15013, 2022.

\bibitem[Nguyen and Tran(2020)]{nguyen2020input}
T.~A. Nguyen and A.~Tran.
\newblock Input-aware dynamic backdoor attack.
\newblock \emph{Advances in Neural Information Processing Systems}, 33:\penalty0 3454--3464, 2020.

\bibitem[Nguyen and Tran(2021)]{nguyen2020wanet}
T.~A. Nguyen and A.~T. Tran.
\newblock Wanet-imperceptible warping-based backdoor attack.
\newblock In \emph{International Conference on Learning Representations}, 2021.

\bibitem[Ren et~al.(2015)Ren, He, Girshick, and Sun]{ren2015faster}
S.~Ren, K.~He, R.~Girshick, and J.~Sun.
\newblock Faster r-cnn: Towards real-time object detection with region proposal networks.
\newblock \emph{Advances in neural information processing systems}, 28, 2015.

\bibitem[Shan et~al.(2020)Shan, Wenger, Wang, Li, Zheng, and Zhao]{shan2020gotta}
S.~Shan, E.~Wenger, B.~Wang, B.~Li, H.~Zheng, and B.~Y. Zhao.
\newblock Gotta catch'em all: Using honeypots to catch adversarial attacks on neural networks.
\newblock In \emph{Proceedings of the 2020 ACM SIGSAC Conference on Computer and Communications Security}, pages 67--83, 2020.

\bibitem[Simonyan and Zisserman(2014)]{simonyan2014very}
K.~Simonyan and A.~Zisserman.
\newblock Very deep convolutional networks for large-scale image recognition.
\newblock \emph{arXiv preprint arXiv:1409.1556}, 2014.

\bibitem[Stallkamp et~al.(2012)Stallkamp, Schlipsing, Salmen, and Igel]{stallkamp2012man}
J.~Stallkamp, M.~Schlipsing, J.~Salmen, and C.~Igel.
\newblock Man vs. computer: Benchmarking machine learning algorithms for traffic sign recognition.
\newblock \emph{Neural networks}, 32:\penalty0 323--332, 2012.

\bibitem[Wang et~al.(2019)Wang, Yao, Shan, Li, Viswanath, Zheng, and Zhao]{wang2019neural}
B.~Wang, Y.~Yao, S.~Shan, H.~Li, B.~Viswanath, H.~Zheng, and B.~Y. Zhao.
\newblock Neural cleanse: Identifying and mitigating backdoor attacks in neural networks.
\newblock In \emph{2019 IEEE Symposium on Security and Privacy (SP)}, pages 707--723. IEEE, 2019.

\bibitem[Wang et~al.(2004)Wang, Bovik, Sheikh, and Simoncelli]{wang2004image}
Z.~Wang, A.~C. Bovik, H.~R. Sheikh, and E.~P. Simoncelli.
\newblock Image quality assessment: from error visibility to structural similarity.
\newblock \emph{IEEE transactions on image processing}, 13\penalty0 (4):\penalty0 600--612, 2004.

\bibitem[Wang et~al.(2022{\natexlab{a}})Wang, Ding, Zhai, and Ma]{wang2022training}
Z.~Wang, H.~Ding, J.~Zhai, and S.~Ma.
\newblock Training with more confidence: Mitigating injected and natural backdoors during training.
\newblock \emph{Advances in Neural Information Processing Systems}, 35:\penalty0 36396--36410, 2022{\natexlab{a}}.

\bibitem[Wang et~al.(2022{\natexlab{b}})Wang, Zhai, and Ma]{wang2022bppattack}
Z.~Wang, J.~Zhai, and S.~Ma.
\newblock Bppattack: Stealthy and efficient trojan attacks against deep neural networks via image quantization and contrastive adversarial learning.
\newblock In \emph{Proceedings of the IEEE/CVF Conference on Computer Vision and Pattern Recognition}, pages 15074--15084, 2022{\natexlab{b}}.

\bibitem[White et~al.(2022)White, Summers, Barnhizer, Barnes, and Snyder]{white2022uniform}
J.~J. White, R.~S. Summers, D.~D. Barnhizer, W.~Barnes, and F.~G. Snyder.
\newblock Uniform commercial code.
\newblock 2022.

\bibitem[Wu and Wang(2021)]{wu2021adversarial}
D.~Wu and Y.~Wang.
\newblock Adversarial neuron pruning purifies backdoored deep models.
\newblock \emph{Advances in Neural Information Processing Systems}, 34:\penalty0 16913--16925, 2021.

\bibitem[Zeng et~al.(2022)Zeng, Chen, Park, Mao, Jin, and Jia]{zengadversarial}
Y.~Zeng, S.~Chen, W.~Park, Z.~Mao, M.~Jin, and R.~Jia.
\newblock Adversarial unlearning of backdoors via implicit hypergradient.
\newblock In \emph{International Conference on Learning Representations}, 2022.

\bibitem[Zhang et~al.(2022)Zhang, Lyu, Wang, Sun, and Sun]{zhanginject}
Z.~Zhang, L.~Lyu, W.~Wang, L.~Sun, and X.~Sun.
\newblock How to inject backdoors with better consistency: Logit anchoring on clean data.
\newblock In \emph{International Conference on Learning Representations}, 2022.

\bibitem[Zheng et~al.(2022)Zheng, Tang, Li, and Liu]{zheng2022data}
R.~Zheng, R.~Tang, J.~Li, and L.~Liu.
\newblock Data-free backdoor removal based on channel lipschitzness.
\newblock In \emph{Computer Vision--ECCV 2022: 17th European Conference, Tel Aviv, Israel, October 23--27, 2022, Proceedings, Part V}, pages 175--191. Springer, 2022.

\bibitem[Zhong et~al.(2022)Zhong, Qian, and Zhang]{nzhong_ijcai}
N.~Zhong, Z.~Qian, and X.~Zhang.
\newblock Imperceptible backdoor attack: From input space to feature representation.
\newblock In \emph{Proceedings of the Thirty-First International Joint Conference on Artificial Intelligence, {IJCAI} 2022, Vienna, Austria, 23-29 July 2022}, pages 1736--1742. ijcai.org, 2022.
\newblock \doi{10.24963/ijcai.2022/242}.
\newblock URL \url{https://doi.org/10.24963/ijcai.2022/242}.

\end{thebibliography}

\end{document}



\begin{frontmatter}


\paperid{207} 


\title{Supplement}





\end{frontmatter}

\section{Attack Effectiveness}

We describe the details of the trigger pattern for baselines as follows. We adopt a random colourful square ($3\times3$, $12 \times 12$ for $32\times32$, $224 \times 224$ size images) in BadNets. In terms of Blend attack, we create poisoned images $x_{p}$ by mixing benign images $x_{b}$ with a randomly selected image $x_{r}$ as $x_{p}= 0.5 * x_{b} + 0.5 * x_{r}$. The trigger pattern of SIG attack is a consecutive vertical stripe. For a benign image whose size is $m\times m$, its trigger can be expressed as $trigger(i,j)=\Delta  sin(2\pi j f/m)$, where $\Delta$ and $f$ mean the intensity and frequency of stripe. We set $\Delta$ and $f$ as 25 and 10, respectively. For WaNet and BppAttack, we create their trigger pattern by their official release with the default setting.

\begin{figure*}[htbp]
   \centering
   \includegraphics[width=6in]{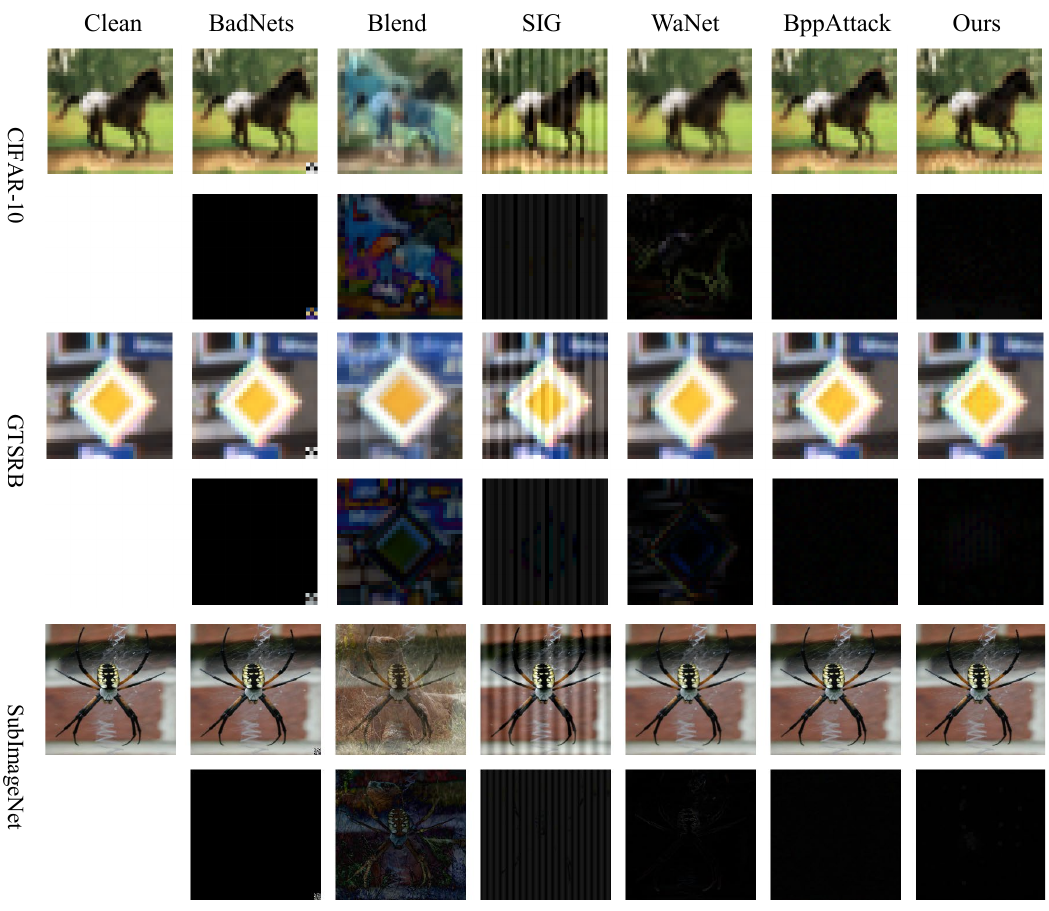}
   \caption{The visualization results of the trigger. The first line denotes clean and poisonous images. The second line denotes the residual between clean and poisonous images.}
   \label{fig_visualization}
\end{figure*}

Table \ref{tab_main}. shows the experiments of VGG networks. The results are consistent with those of ResNet-18 shown in the manuscript. Our approach achieves satisfactory attack effectiveness, fidelity, and revocability compared with baselines. Fig. \ref{fig_visualization} shows the visualization of three datasets and our trigger pattern is invisible to human inspection.

\begin{figure*}[] 
   \centering
   \includegraphics[width=6in]{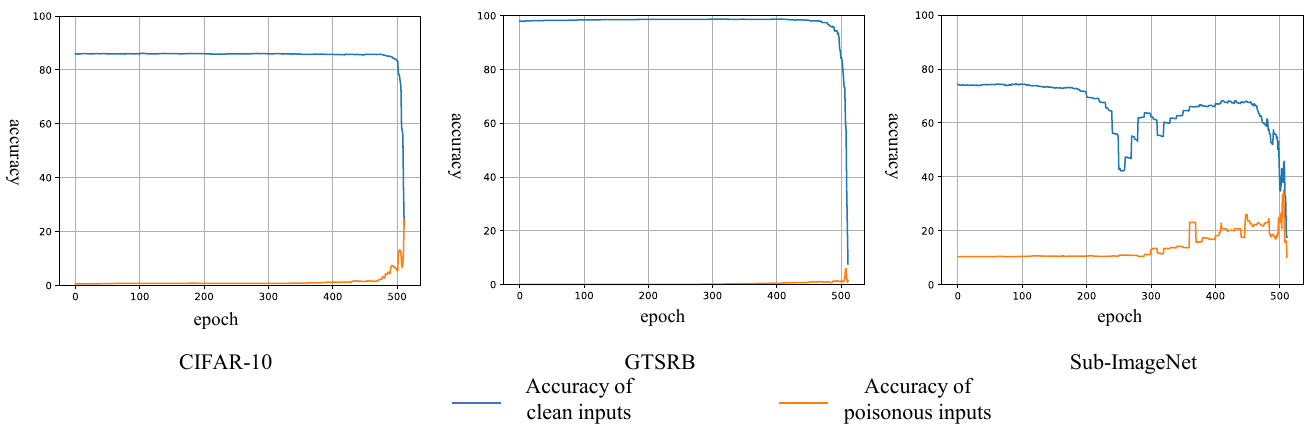}
   \caption{The experimental results of fine-pruning defence. (ResNet-18)}
   \label{fig_robust}
\end{figure*}

\begin{figure*}[t]
   \centering
   \includegraphics[width=6 in]{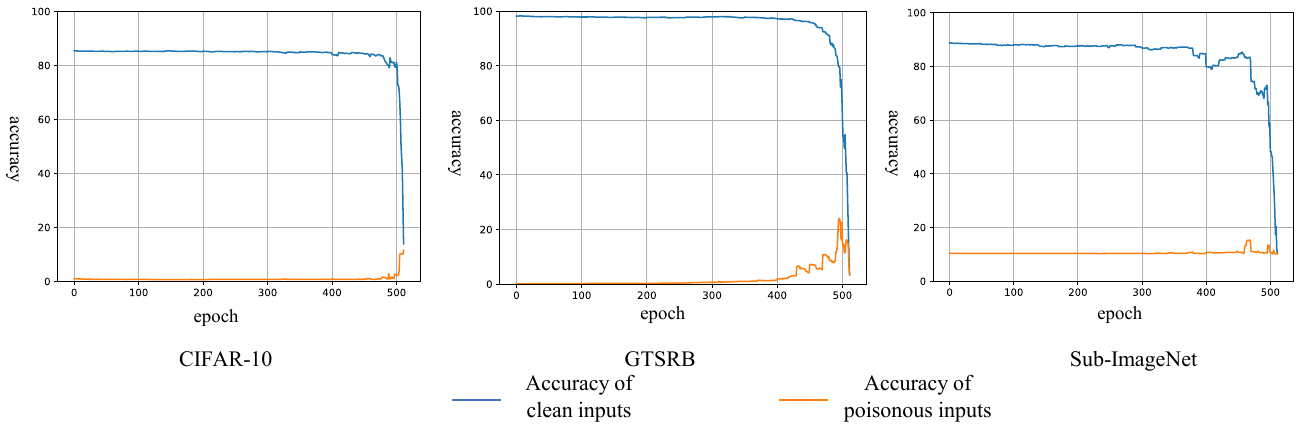}
   \caption{The experimental results of fine-pruning defence (VGG).}
   \label{fig_pruning}
\end{figure*}

Due to the limitation of the page, we do not show the attack robustness over VGG network in the manuscript. Table \ref{tab_NAD_FT} shows the experimental results including fine-tuning, fine-pruning, and NAD defence for VGG network. The experimental results are also consistent with those of ResNet-18. The details of the accuracy (both clean and poisonous inputs) with regard to the number of pruned neurons can be found in Fig. \ref{fig_pruning} and Fig. \ref{fig_robust}. These backdoor purification defences are ineffective to our approach.

Hyperparameter confidence $c$ is used to balance two terms of the following equation,
\begin{eqnarray}
   L_{bd} = \frac{1}{{|D_c|}}\sum\limits_{(x,y) \in D_c} \ell  \left( {{f_\theta }(x),y} \right) \\ - \frac{1}{{|D_p|}}\sum\limits_{(x,y) \in D_p} max( \ell   \left( {{f_\theta }(x+\delta),y),c} \right). \label{e1}
\end{eqnarray}
As described in the manuscript. The classifier $f_\theta(\cdot)$ easily learns how to give wrong results for poisonous inputs but it is hard to learn how to give correct results for clean inputs. Therefore, we adopt $max(\cdot)$  and hyperparameter confidence $c$ to restrict the second term. In this part, we vary $c$ from $-0.5*log(1/N)$ to $+\infty$ as shown in Table \ref{tab_c}. It is equivalent to removing the restriction of $max(\cdot)$ when $c=+\infty$. From Table \ref{tab_c}, we find that $c=-1.1*log(1/N)$ is a suitable selection. The training of the classifier will collapse without the restriction of $c$. However, too small $c$  decreases the attack effectiveness.

The regularization $L_R$ for mask matrices $M_i$ has an impact on the performance of our approach. We vary $\alpha$ used to control the balance of $L_R$ from 0 to 100 to demonstrate the usage of $L_R$ over Sub-ImageNet dataset. The experimental results in Table \ref{tab_L_R} show that regularization $L_R$ significantly improves the accuracy of clean inputs for compromised classifiers. 

We also consider the uniqueness of the mask matrices. In our approach, the mask matrices are generated by minimizing equation 7 which requires clean and poisonous images. We suppose buyers, who know our approach in advance, have a few clean images and a model containing a revocable backdoor. They try to generate spurious mask matrices by minimizing equation 7 without poisonous images. Our experiment shows that spurious mask matrices cannot eliminate our backdoor. The attack success rate is almost unchanged in all cases. The poisonous images are crucial to generate the mask matrices. Buyers cannot access poisonous images resulting in the failure of generating effective mask matrices.

\begin{table}[htbp]
  \centering
  
    \begin{tabular}{ccccc}
    \toprule
          & \multicolumn{2}{c}{Effectiveness } & \multicolumn{2}{c}{Revocability} \\
    $\alpha$  & $Acc_{BD}$-C & $Acc_{BD}$-P  & $Acc_{BD}$-C  & $Acc_{BD}$-P \\
    \hline
    0     & 70.42 & 10.09 & 73.42 & 72.78 \\
    10    & 72.98 & 10.36 & 73.05 & 71.47 \\
    100   & 74.06  & 10.43  & 74.33  & 71.33  \\
    \bottomrule
    \end{tabular}%
    \caption{The ablaiton studies of $L_R$ over Sub-ImageNet and ResNet-18.}
  \label{tab_L_R}%
\end{table}%

\begin{table*}[htbp]
  \centering
  
    \begin{tabular}{cccccc}
    \toprule
    Dataset &       & \multicolumn{2}{c}{Effectiveness } & \multicolumn{2}{c}{Revocability} \\
    $Acc_{Clean}$ & Attack & $Acc_{BD}$-C & $Acc_{BD}$-P & $Acc_{BD}$-C & $Acc_{BD}$-P \\
    \hline
          & BadNets & 83.57  & 10.01  & 79.99  & 79.64  \\
          & Blend & 83.44  & 10.02  & 78.32  & 66.84  \\
    CIFAR-10  & SIG   & 83.52  & 10.04  & 78.40  & 76.43  \\
    84.72 & LSB   & 79.12  & 79.23  & 71.61  & 71.58  \\
          & WaNet & 79.86  & 36.47  & 76.60  & 75.32  \\
          & BppAttack & 79.30  & 79.02  & 72.22  & 72.02  \\
          & Ours  & 85.17  & 1.00  & 85.29  & 84.35  \\
          \hline
          & BadNets & 97.27  & 0.00  & 95.46  & 95.53  \\
          & Blend & 96.91  & 0.00  & 94.85  & 92.07  \\
    GTSRB & SIG   & 97.05  & 0.00  & 94.80  & 94.16  \\
    97.87 & LSB   & 95.84  & 9.11  & 95.84  & 95.73  \\
          & WaNet & 94.44  & 12.02  & 95.84  & 95.96  \\
          & BppAttack & 86.90  & 4.02  & 94.11  & 94.03  \\
          & Ours  & 98.19  & 0.28  & 98.16  & 97.71  \\
          \hline
          & BadNets & 88.19  & 10.33  & 81.46  & 81.29  \\
          & Blend & 88.32  & 10.80  & 80.32  & 76.88  \\
    Sub-ImageNet & SIG   & 87.72  & 10.09  & 81.70  & 81.12  \\
    90.11 & LSB   & 84.76  & 84.59  & 72.51  & 72.61  \\
          & WaNet & 83.18  & 80.38  & 72.95  & 73.25  \\
          & BppAttack & 84.15  & 83.55  & 74.02  & 73.18  \\
          & Ours  & 88.43  & 10.40  & 88.49  & 87.79  \\
          \bottomrule
    \end{tabular}%
    \caption{The experimental comparison (VGG) between baselines and ours. $Acc_{Clean}$, $Acc_{BD}$-C, and $Acc_{BD}$-P denote the accuracy of the clean classifier, the accuracy of the compromised classifier for clean inputs, and the accuracy of the compromised classifier for poisonous inputs, respectively. Effectiveness and revocability denote the compromised classifier and compromised classifier with mask matrices, respectively. }
  \label{tab_main}%
\end{table*}%

\begin{table*}[]
  \centering
  
    \begin{tabular}{ccccccccc}
    \toprule
          & \multicolumn{2}{c}{No defence} & \multicolumn{2}{c}{NAD} & \multicolumn{2}{c}{Fine-tuning} & \multicolumn{2}{c}{Fine-pruning} \\
          \hline
    Dataset & $Acc_{BD}$-C & $Acc_{BD}$-P & $Acc_{BD}$-C & $Acc_{BD}$-P & $Acc_{BD}$-C & $Acc_{BD}$-P & $Acc_{BD}$-C & $Acc_{BD}$-P \\
    \hline
    CIFAR-10 & 85.17  & 1.00  & 71.14 & 66.44 & 85.28 & 0.87  &  84.12 & 0.94 \\
    GTSRB & 98.19  & 0.28  & 36.04 & 34.08 & 98.15 & 0.17  & 97.13  &  3.06 \\
    SubImageNet & 88.43  & 10.40  & 88.76 & 10.33 & 88.46 & 10.30   & 87.76 & 10.41 \\
    \bottomrule
    \end{tabular}%
    \caption{The experimental results (VGG) of NAD defence and fine-tuning.}
  \label{tab_NAD_FT}%
\end{table*}%

\begin{table*}[htbp]
  \centering
  
    \begin{tabular}{ccccc}
    \toprule
          & \multicolumn{2}{c}{Effectiveness } & \multicolumn{2}{c}{Revocability} \\
    $c$ & $Acc_{BD}$-C & $Acc_{BD}$-P  & $Acc_{BD}$-C  & $Acc_{BD}$-P \\
    \hline
    $-0.5*\log(1/N)$   & 74.13 & 72.41 & 74.16 & 72.41 \\
    $-1.1*\log(1/N)$   & 74.06  & 10.43  & 74.33  & 71.33  \\
    $+\infty$  & 10.77  & 10.09  & 10.06  & 10.09  \\
    \bottomrule
    \end{tabular}%
    \caption{The ablaiton studies of $c$ over Sub-ImageNet and ResNet-18.}
  \label{tab_c}%
\end{table*}%





